\documentclass[%
 reprint,
 amsmath,amssymb,
pra,
]{revtex4-1}
\usepackage{color}
\usepackage{epstopdf}
\usepackage{graphicx}
\usepackage{dcolumn}
\usepackage{bm}
\usepackage{amssymb}   
\usepackage{amsmath}
\usepackage{mathrsfs}
\usepackage{braket}
\usepackage{hyperref}

\begin{document}

\preprint{APS/123-QED}


\title {Enhancing quantum correlations and entanglement in an optomechanical system via cross-Kerr nonlinearity}

\author{Subhadeep Chakraborty}
 \email{c.subhadeep@iitg.ernet.in}
\author{Amarendra K. Sarma}%
 \email{aksarma@iitg.ernet.in}
\affiliation{%
 Department of Physics, Indian Institute of Technology  Guwahati, Guwahati-781039, Assam, India}


%

\date{\today}

\begin{abstract}
In this work, we theoretically study the quantum correlations present in an optomechanical system by invoking an additional cross-Kerr coupling between the optical and mechanical mode. Under experimentally achievable conditions,  we first show that a significant enhancement of the steady-state entanglement could be achieved at a considerably lower driving power, which is also extremely robust with respect to system parameters and environmental temperature. Then, we employ Gaussian quantum discord as a witness of the genuine quantumness of the correlation present in the system and discuss its dependence on the cross-Kerr nonlinearity.
\begin{description}
\item[PACS numbers]
\verb|42.50.Wk, 03.65.Ud, 07.10.Cm, 42.50.Lc|
\end{description}
\end{abstract}

\pacs{Valid PACS appear here}
\maketitle


\section{\label{sec:level1}INTRODUCTION}
Entanglement \cite{1,2} is one of the most intriguing feature of quantum mechanics, having myriads of potential applications in quantum information and communication \cite{3}. So far, preparation and manipulation of entanglement has been successfully demonstrated in microscopic systems, such as atoms , photons, ions \cite{4,5,6} etc. However, the validity of entanglement in the macroscopic realm is still a debatable fact. In this regard, optomechanical system \cite{7,8} where a macroscopic mechanical motion interacts with an optical field has emerged as a promising platform to realize entanglement at a macroscopic level. Several studies have been reported to generate entanglement between optical and mechanical mode or two optical modes or two mechanical modes \cite{9,10,11,12,13,14,15,16,17,18,19,20,21,22,23,24,25,26,27,28,29,30,31,32,33,34}. These studies mostly explore entanglement in the steady-state regime of the system where the strength of the entanglement is strictly limited by the stability conditions. In particular, in this regime entanglement becomes maximum only when the system is driven close to the instability threshold, which demands a strong multi-photon optomechanical coupling or high external laser driving \cite{11}.

Besides the quantum entanglement, quantum discord is another measure for the quantumness of the correlations present in a quantum state. In the seminal paper \cite{35d}, Zurek introduced discord as the mismatch between the two classically equivalent description of mutual information and showed that it can even exists for separable states (which are usually referred to have classical correlation). However, their study was limited in the finite dimensional system, which was further extended in the realm of continuous-variables (CV) in Refs. \cite{36d,37d}. In recent years, quantum discord has received much attention in optomechanical systems to study non-classical correlations in it \cite{38d}. In particular, it has been both theoretically and experimentally demonstrated that pre-availability of such non-classical correlation can activate entanglement, which is also more robust against temperature and thermal noise \cite{39d,40d}.

Very recently, optomechanical systems coupled to a two-level system or a nonlinear inductive element (single-Cooper pair transistor) have shown significant enhancement of the single-photon radiation-pressure coupling \cite{35,36}. More importantly, it has been theoretically deduced that such interaction induces an additional cross-Kerr type coupling between the optical and mechanical mode. In nonlinear optics, cross-Kerr effect describes the change in refractive index of one electromagnetic mode by the intensity of another. It has been extensively studied in the context of quantum information processing, such as non-demolition photon number detection \cite{37}, C-NOT gate \cite{38} and, discrete \cite{39} and continuous-variable entanglement concentration \cite{40,41,42}. Concerning optomechanical system, the cross-Kerr coupling between the optical and mechanical mode, involves the change in the refractive index of the optical mode depending on the resonators phonon number \cite{43,44}. Recently, Ref. \cite{45} has showed that this cross-Kerr coupling gives rise to a frequency shift in mechanical and optical mode, with the shift depending on the cavity photon number and mechanical phonon number. Furthermore, Ref. \cite{46} studied the effect of cross-Kerr nonlinearity on the stability of the optomechanical system. Along this line, Ref. \cite{47} showed that in presence of an additional cross-Kerr coupling, the steady-state response of the mechanical resonator becomes a nonmonotonous function of cavity photon number and the bistable behavior of the mean cavity photon can be turned into a tristable behavior.

Motivated by these works, we study the combined effect of radiation pressure and cross-Kerr coupling on the quantum correlations present in an optomechanical system. We start by analyzing the effect of cross-Kerr coupling on the steady-state behavior and the stability conditions of the optomechanical system. Then, we focus on the optomechanical entanglement and find the influence of cross-Kerr coupling on it. Finally, we employ Gaussian quantum discord as an additional measure of quantum correlation beyond entanglement, and discuss its dependence on the cross-Kerr coupling.
\begin {figure}[t]
\begin {center}
\includegraphics [width =6cm]{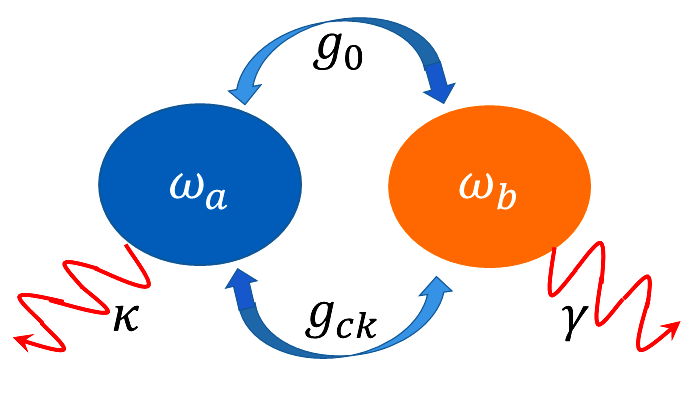}
\caption{\label {fig1}(Color online) Schematic diagram of the considered optomechanical system. An optical mode (with frequency $\omega_a$) couples the mechanical mode (with frequency $\omega_b$) via the radiation-pressure coupling $g_0$ and an additional cross-Kerr coupling $g_{ck}$. Here, $\kappa$ is the optical decay rate and $\gamma$ is the mechanical damping rate.}
\end{center}
\end{figure}

\section{\label{sec:level2}MODEL AND EQUATIONS}
As depicted in Fig. \ref{fig1}, we consider an optomechanical system which consists of an optical and a mechanical mode, respectively, with frequency $\omega_a$ and $\omega_b$. As usual, these two modes are coupled via the generic radiation pressure coupling $g_0$. In addition, here we consider an extra cross-Kerr type coupling $g_{ck}$ between the optical and the mechanical mode, generated by a two-level system or a superconducting charge qubit \cite{35,36}. Also, the whole system is driven by an external laser of frequency $\omega_l$. With this consideration, the complete Hamiltonian of the optomechanical system in a rotating frame of the laser reads ($\hbar=1$):
\begin{gather}\label{eq:1}
H=\Delta_0a^\dagger a+\omega_b b^\dagger b-g_0a^\dagger a(b^\dagger+b)-g_{ck}a^\dagger ab^\dagger b\\\nonumber
+iE_0(a^\dagger-a).
\end{gather}
Here, $a$ ($a^\dagger$) and $b$ ($b^\dagger$) are, respectively, the annihilation (creation) operators of the optical and mechanical mode, $\Delta_0=\omega_a-\omega_l$ is the optical detuning and $E_0$ is the driving amplitude, related to driving power $P$ and optical decay rate $\kappa$ by $|E_0|=\sqrt{2P\kappa/\hbar\omega_l}$.

In addition, the system dynamics also includes fluctuation-dissipation processes affecting both the optical and the mechanical mode. Starting from Hamiltonian \eqref{eq:1}, the time evolution of the optomechanical system is given by the following nonlinear quantum Langevin equations:
\begin{subequations}\label{eq:2}
\begin{gather}
\dot{a}=-(i\Delta_0+\kappa)a+ig_0a(b^\dagger+b)+ig_{ck}ab^\dagger b+E_0\\\nonumber
+\sqrt{2\kappa}a_{in},\\
\dot{b}=-(i\omega_b+\gamma)b+ig_0a^\dagger a+ig_{ck}a^\dagger ab+\sqrt{2\gamma}b_{in},
\end{gather}
\end{subequations}
where $\gamma$ is the mechanical damping rate, and $a_{in}$ and $b_{in}$ are the corresponding zero-mean input Gaussian noises, with nonzero correlation functions \cite{48}, given by:
\begin{subequations}\label{eq:3}
\begin{gather}
\langle a_{in}(t)a^\dagger _{in}(t^\prime)\rangle=\delta(t-t^\prime),\\
\langle b^\dagger _{in}(t)b_{in}(t^\prime)\rangle=n_{th}\delta(t-t^\prime),\\
\langle b_{in}(t)b^\dagger _{in}(t^\prime)\rangle=(n_{th}+1)\delta(t-t^\prime).
\end{gather}
\end{subequations}
Here, $n_{th}$ is the mean thermal phonon number, related to Boltzmann constant $K_B$ and environmental temperature $T$ by: $n_{th}=\left[\mathrm{ exp}\left(\frac{\hbar \omega_b}{K_BT}\right)-1\right]^{-1}$.

Under intense laser driving, we now expand each Heisenberg operators as a sum of its semi-classical steady-state value plus an additional small fluctuation operator with zero-mean value: $a=\alpha+\delta a$ and $b=\beta+\delta b$. The steady-state values are given by the following nonlinear equations:
\begin{subequations}
\begin{gather}
(i\tilde{\Delta}+\kappa)\alpha-E_0=0,\label{eq:4(a)}\\
(i\tilde{\omega_b}+\gamma)\beta-ig_0|\alpha|^2=0\label{eq:4(b)},
\end{gather}
\end{subequations}
where $\tilde{\Delta}=\Delta_0-g_0(\beta^*+\beta)-g_{ck}|\beta|^2$ is the effective optical detuning, modified owing to the both radiation pressure and cross-Kerr coupling, and, $\tilde{\omega_b}=\omega_b-g_{ck}|\alpha|^2$ is the effective mechanical frequency, modified owing to the cross-Kerr interaction. As for the fluctuations, their dynamics is given by the linearized quantum Langevin equations \cite{10}:
\begin{subequations}\label{eq:5}
\begin{gather}
\dot{\delta a}=-(i\tilde{\Delta}+\kappa)\delta a+i\frac{G}{2}(\delta b^\dagger+\delta b)+\sqrt{2\kappa}a_{in},\\
\dot{\delta b}=-(i\tilde{\omega_b}+\gamma)\delta b+i\frac{G}{2}(\delta a^\dagger+\delta a)+\sqrt{2\gamma}b_{in},
\end{gather}
\end{subequations}
where $G=2g|\alpha|$ is the effective optomechanical coupling strength with modified $g=g_0+g_{ck}\beta$. Note that, in the above calculations the phase reference of the optical field is chosen such that $\alpha$ is real.

Next, we introduce the dimensionless quadrature operators, respectively, for the optical and the mechanical mode: $\delta X\equiv\frac{\left(\delta a+\delta a^\dagger\right)}{\sqrt2}$, $\delta Y\equiv\frac{\left(\delta a-\delta a^\dagger\right)}{i\sqrt2}$ and $\delta Q\equiv\frac{\left(\delta b+\delta b^\dagger\right)}{\sqrt2}$, $\delta P\equiv\frac{\left(\delta b-\delta b^\dagger\right)}{i\sqrt2}$, and, similarly for the corresponding Hermitian input noise operators: $X_{in}\equiv\frac{\left(a_{in}+a_{in}^\dagger\right)}{\sqrt2}$, $Y_{in}\equiv\frac{\left(a_{in}-a_{in}^\dagger\right)}{i\sqrt2}$ and $Q_{in}\equiv\frac{\left(b_{in}+b_{in}^\dagger\right)}{\sqrt2}$, $P_{in}\equiv\frac{\left(b_{in}-b_{in}^\dagger\right)}{i\sqrt2}$. With the above definitions, we express Eq. \eqref{eq:5} in a more compact form:
\begin{align}\label{eq:6}
\dot u(t)=Au(t)+n(t),
\end{align}
where $u^T(t)=\left(\delta Q(t),\delta P(t),\delta X(t),\delta Y(t)\right)$ is the vector of continuous variable (CV) fluctuation operators, $A$ is the drift matrix:
\begin{align}\label{eq:7}
A=\left(
\setlength\arraycolsep{6pt}
  \begin{array}{cccc}
    -\gamma & \tilde{\omega_b} & 0 & 0 \\[4 pt]
    -\tilde{\omega_b} & -\gamma & G & 0 \\[4 pt]
    0 & 0 & -\kappa & \tilde{\Delta} \\[4 pt]
    G & 0 & -\tilde{\Delta} & -\kappa \\[4 pt]
  \end{array}
\right),
\end{align}
and $n^T(t)=\left(\sqrt{2\gamma}Q_{in}(t),\sqrt{2\gamma}P_{in}(t),\sqrt{2\kappa}X_{in}(t),\sqrt{2\kappa}Y_{in}(t)\right)$ is the vector of corresponding noises. A formal solution of Eq. \eqref{eq:6} is given by:
\begin{align}\label{eq:8}
u(t)=M(t)u(0)+\int_0^t ds M(s)n(t-s),
\end{align}
where $M(t)=e^{AT}$. The system is stable and reaches its steady-state when all the eigenvalues of the drift matrix $A$ have negative real parts. These stability conditions are derived by applying the Routh-Hurwitz criteria \cite{49},which is given in terms of the system parameters by the following two nontrivial equations:
\begin{subequations}\label{eq:9}
\begin{gather}
4\gamma\kappa[{\tilde{\Delta}}^4+2{\tilde{\Delta}}^2(\gamma^2+\kappa^2-{\tilde{\omega_b}}^2)+(\gamma^2+\kappa^2+{\tilde{\omega_b}}^2)^2\\\nonumber
+4\gamma\kappa(\tilde{\Delta}+\kappa+\gamma^2+\kappa^2+{\tilde{\omega_b}}^2)]+4G^2\tilde{\Delta}\tilde{\omega_b}(\gamma+\kappa)^2>0,\\
({\tilde{\Delta}}^2+\kappa^2)({\tilde{\omega_b}}^2+\gamma^2)-G^2\tilde{\Delta}\tilde{\omega_b}>0.
\end{gather}
\end{subequations}
Note that, in the following work we will strictly restrict to red-detuned driving ($\tilde{\Delta}>0$), for which the first condition is always satisfied, only the second condition is relevant.
\begin {figure}[t]
\begin {center}
\includegraphics [width =7cm]{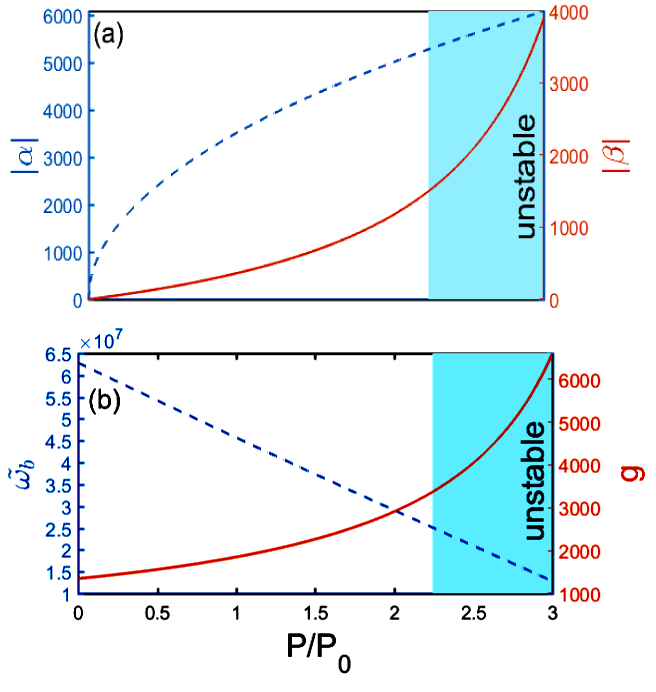}
\caption{\label {fig2}(Color online) (a) The steady-state amplitudes $|\alpha|$ (blue dashed) and $|\beta|$ (red solid), and, (b) the modified mechanical frequency $\tilde{\omega_b}$ (blue dashed) and coupling strength $g$ (red solid) versus the driving power $P/P_0$. The shaded (cyan) region corresponds to the unstable phase of the optomechanical system. The other parameters are fixed at: $g_{ck}=10^{-3}g_0$, $\kappa/\omega_b=0.4$ and $\Delta_0/\omega_b=0.5$.}
\end{center}
\end{figure}

To numerically illustrate the effect of the cross-Kerr coupling on the steady-state behavior of the optomechanical system, we consider a set of experimentally accessible parameters: $\omega_a/2\pi=370$ THz, $\omega_b/2\pi=10$ MHz, $g_0=1.347$ KHz, $\gamma/2\pi=100$ Hz, $P_0=0.1$ mW and $n_{th}=100$. In Fig. 2(a), we plot the steady-state amplitudes of the optical and mechanical mode along with the stability range of the system (with $g_{ck}=10^{-3}g_0$), as a function of the driving power. It is clear that for the chosen value of the parameters, both these amplitudes satisfy $|\alpha|, |\beta|\gg1$ which ensures the validity of our linearization assumption. We also note that unlike pure radiation-pressure coupling, now $|\beta|$ changes drastically with increasing power, as could be explained from Eq. \eqref{eq:4(b)}.  In Fig. (2b) we plot  $\tilde{\omega_b}$ and $g$ against the driving power. We can see that with increasing power $\tilde{\omega_b}$ decreases and $g$ increases. In particular, at a driving power $P=2.23P_0$ (just before the instability), we get $\tilde{\omega_b}\approx0.4\omega_b$ and $g\approx2.5g_0$. Therefore, in presence of the cross-Kerr coupling, owing to this significant change in the effective mechanical frequency and (single photon) coupling strength, the system becomes unstable at a considerably lower driving power.It is worthwhile to note that in Ref. \cite{45}, Raphael et al. have shown that the frequencies of the mechanical resonator and the optical cavity gets shifted owing to the cross-Kerr effect and the shift depends on the number of photons in the cavity and phonons in the resonators. On the other hand, in Ref. \cite{47}, Wei et al. show that the mean phonon number, which increases monotonically with increase in the photon number without cross-Kerr effect, changes drastically with introduction of the cross-Kerr effect. Our analysis, as could be seen from Eq.(4) and Fig. 2, agrees well with these observations. However, in this work, the parameter regime where the system becomes unstable is of prime concern to us owing to its significance in attaining high entanglement.
\section{\label{sec:level3}OPTOMECHANICAL ENTANGLEMENT}
\begin {figure}[b]
\begin {center}
\includegraphics [width =7cm]{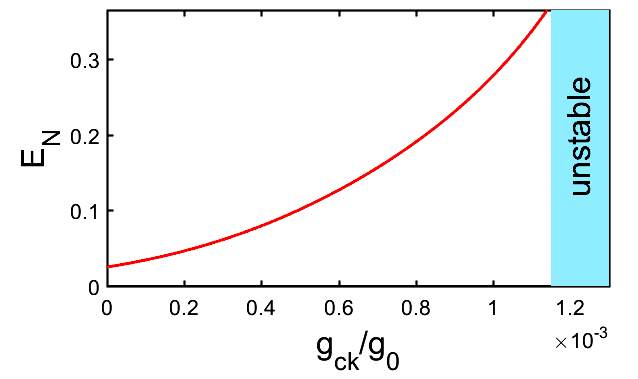}
\caption{\label {fig3}(Color online) $E_N$ as a function of the normalized cross-Kerr coupling strength $g_{ck}/g_0$, at a fixed $P/P_0=2.0$, $\Delta_0/\omega_b=0.5$ and $\kappa/\omega_b=0.4$.}
\end{center}
\end{figure}
Since the dynamics is linearized and the initial state of the system is Gaussian in nature, the steady-state for the quantum fluctuations is simply a zero-mean Gaussian bipartite state, which is fully characterized by its $4\times4$ correlation matrix (CM):
\begin{align}\label{eq:10}
V_{ij}=\left(\langle u_i(\infty)u_j(\infty)+u_j(\infty)u_i(\infty)\rangle\right)/2.
\end{align}
Here, $u^T(\infty)=\left(\delta Q(\infty),\delta P(\infty),\delta X(\infty),\delta Y(\infty)\right)$ is the vector of CV fluctuation operators in the steady-state $(t\rightarrow\infty)$. When the system is stable, performing a substitution of Eq. \eqref{eq:8} in the definition of CM, we get
\begin{align}\label{eq:11}
V_{ij}=\sum_{k,l}\int_0^\infty ds\int_0^\infty ds^\prime M_{ik}(s)M_{jl}(s^\prime)\Phi_{kl}(s-s^\prime),
\end{align}
where $\Phi_{kl}(s-s^\prime)=(\langle n_k(s)n_l(s^\prime)+n_l(s^\prime)n_k(s)\rangle)/2$ is the matrix of the stationary noise correlations. Using Eq. \eqref{eq:3}, $\Phi_{kl}(s-s^\prime)$ further simplifies to $\Phi_{kl}(s-s^\prime)=\textit{D}_{kl}\delta(s-s^\prime)$, where $\textit{D}$ is a diagonal matrix, given by:
\begin{align}\label{eq:12}
\textit{D}=\mathrm{Diag}\left[\gamma(2n_{th}+1),\gamma(2n_{th}+1),\kappa,\kappa\right].
\end{align}
\begin {figure*}[t]
\begin {center}
\includegraphics [width =15cm]{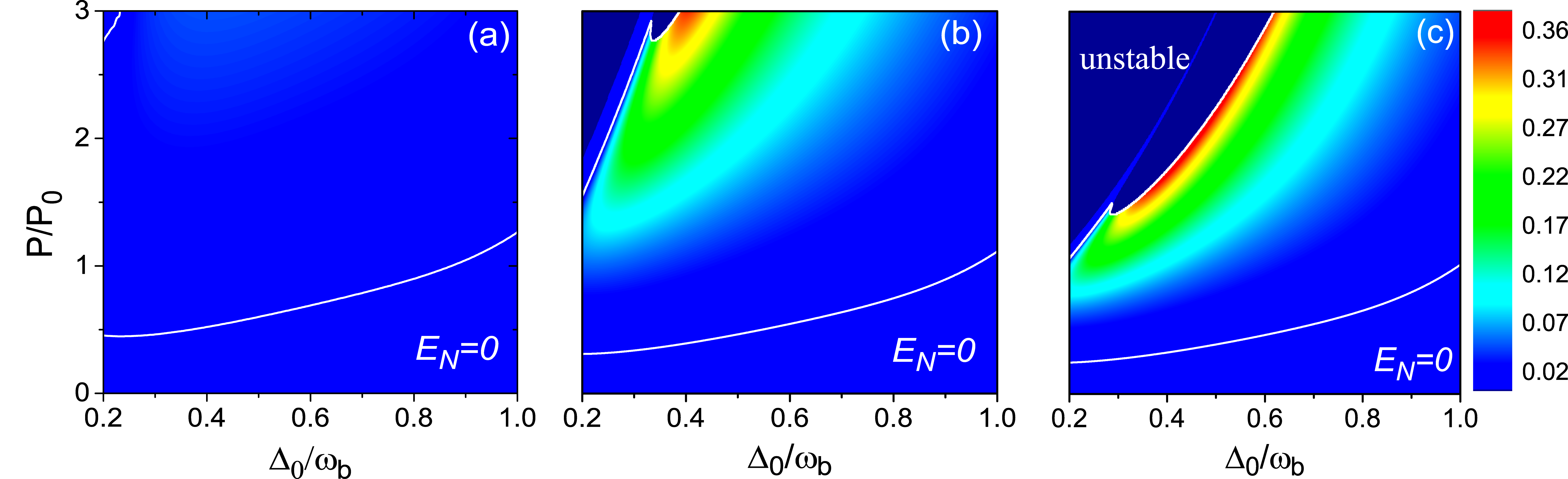}
\begin{flushleft}
\caption{\label {fig4}(Color online) Contour plot of $E_N$ versus the normalized detuning $\Delta_0/\omega_b$ and driving power $P/P_0$, for $g_{ck}=0$ (a) (no cross-Kerr coupling), $g_{ck}=0.5\times 10^{-3} g_0$ (b), and $g_{ck}=1.0\times 10^{-3} g_0$ (c). The optical decay rate is fixed at $\kappa/\omega_b=0.4$.}
\end{flushleft}
\end{center}
\end{figure*}
\begin {figure*}[!htb]
\begin {center}
\includegraphics [width =15cm]{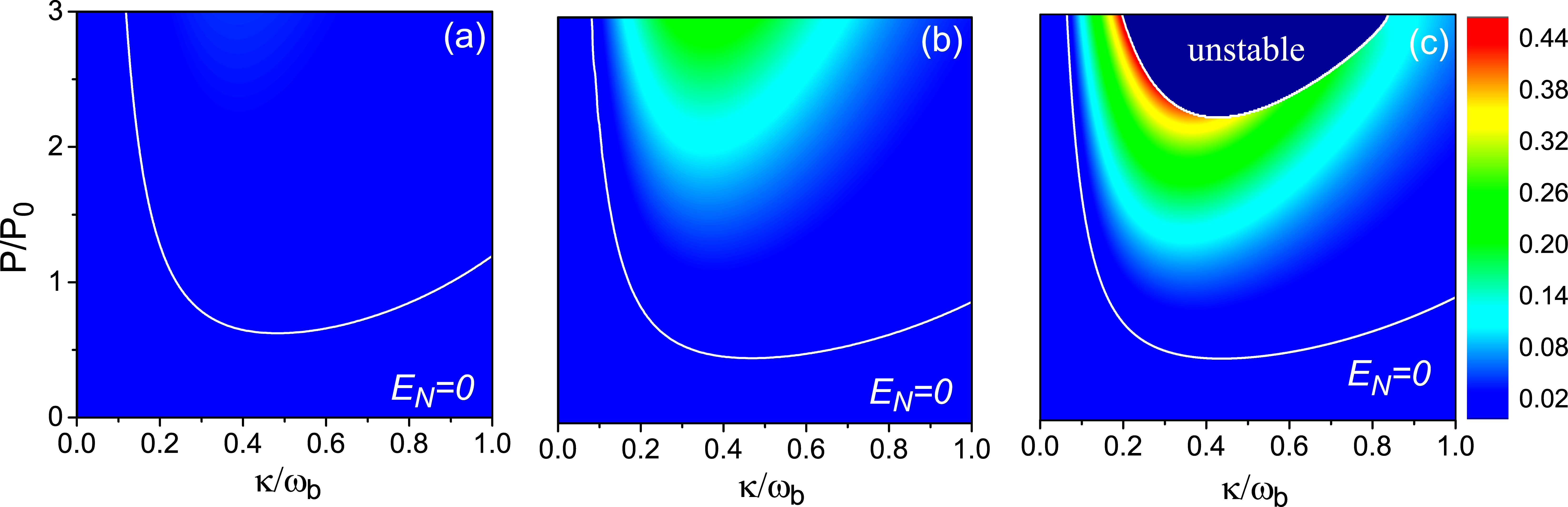}
\begin{flushleft}
\caption{\label {fig5}(Color online) Contour plot of $E_N$ versus the normalized optical decay rate $\kappa/\omega_b$ and normalized driving power $P/P_0$, for $g_{ck}=0$ (a) (no cross-Kerr coupling), $g_{ck}=0.5\times 10^{-3} g_0$ (b), and $g_{ck}=1.0\times 10^{-3} g_0$ (c). The optical detuning  is fixed at $\Delta_0/\omega_b=0.5$.}
\end{flushleft}
\end{center}
\end{figure*}
Therefore, Eq. \eqref{eq:11} becomes $V=\int_0^\infty dsM(s)DM(s)^T$, which, when the stability conditions are satisfied ($M(\infty)=0$) is equivalent to the following Lyapunov equation for $V$:
\begin{align}\label{eq:13}
AV+VA^T=-D.
\end{align}

We now express the above correlation matrix $V$ in a $2\times2$ block from:
\begin{align}\label{eq:14}
V=\left(
  \begin{array}{cc}
    V_m & V_{corr} \\
    V_{corr}^T & V_o \\
  \end{array}
\right),
\end{align}
where $V_m$, $V_o$ and $V_{corr}$, respectively  corresponds to the mechanical mode, the optical mode and the optomechanical correlation between them, and define four symplectic local invariants:
\begin{align}\label{eq:15}
I_1=\mathrm{det}V_m, \quad I_2=\mathrm{det}V_o, \\\nonumber I_3=\mathrm{det}V_{corr}, \quad I_4=\mathrm{det}V.
\end{align}
Then, the degree of quantum entanglement between the optical and mechanical mode can be assessed by calculating the so-called logarithmic negativity $E_N$ \cite{50,51}, defined as:
\begin{align}\label{eq:16}
E_N=\mathrm{max}\left[0,-\ln2\tilde{\nu}^-\right].
\end{align}
where, $\tilde{\nu}^-\equiv2^{-1/2}\left[\tilde{\Delta}(V)-\sqrt{\tilde{\Delta}(V)^2-4I_4}\right]^{1/2}$ is the smallest symplectic eigenvalue of the partial transpose of $V$ with $\tilde{\Delta}(V)\equiv I_1+I_2-2I_3$. Therefore, a Gaussian state is entangled ($E_N>0$) if and only if $\tilde{\nu}^-<1/2$ which is equivalent to Simon's necessary and sufficient nonpositive partial transpose criteria \cite{52}.

First, we focus on the steady-state optomechanical entanglement and the influence of cross-Kerr coupling on it. In Fig. \ref{fig3}, we plot $E_N$ as a function of the normalized cross-Kerr coupling strength $g_{ck}/g_0$, for a fixed driving power and optical detuning. We can see that with increase in the cross-Kerr coupling, the degree of entanglement increases and becomes maximum just before the instability. In particular, for the pure radiation-pressure coupling case ($g_{ck}=0$) we have $E_N=0.025$. On the other hand, in the presence of cross-Kerr coupling, with coupling strength $g_{ck}=10^{-3}g_0$, we obtain $E_N=0.27$. Therefore, by invoking cross-Kerr coupling  on a generic optomechanical system, we can significantly enhance the degree of the steady-state optomechanical entanglement.

To further probe into entanglement properties and its dependence on the system parameters, we plot in Fig.\ref{fig4}, $E_N$ as a function of the normalized detuning $\Delta_0/\omega_b$ and the normalized driving power $P/P_0$, for $g_{ck}=0$ (a) (no cross-Kerr coupling), $g_{ck}=0.5\times 10^{-3} g_0$ (b), and $g_{ck}=1.0\times 10^{-3} g_0$ (c). We observe that with increase in the cross-Kerr coupling strength, the degree of entanglement increases, however, the parameter region in which the steady-state is entangled significantly narrows. In fact, with increasing cross-Kerr coupling strength, as discussed in earlier section, the system becomes unstable at a lower driving power, which leads to the significant enhancement of steady-state optomechanical entanglement near the instability threshold.

Fig.\ref{fig5}, depicts the same $E_N$ as a function of the normalized optical decay rate $\kappa/\omega_b$ and normalized driving power $P/P_0$, for $g_{ck}=0$ (a) (no cross-Kerr coupling), $g_{ck}=0.5\times 10^{-3} g_0$ (b), and $g_{ck}=1.0\times 10^{-3} g_0$ (c). Here also, we find the enhancement of the steady-state optomechanical with increasing cross-Kerr coupling strength. However, now the maximum entanglement only occurs for $g_{ck}=1\times 10^{-3}g_0$ (Fig. 5(c)) at higher driving power and lower optical decay rate. It is also important to note that even in the presence of cross-Kerr coupling, we find significant entanglement only in the resolved sideband region, i.e. for $\kappa/\omega_b<1$.
\begin {figure}[!htb]
\begin {center}
\includegraphics [height=4.7cm,width =7cm]{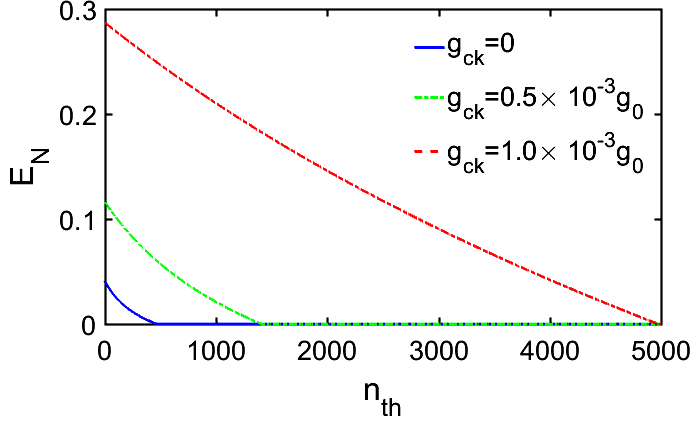}
\begin{flushleft}
\caption{\label {fig6}(Color online) $E_N$ versus the mean thermal phonon number for different cross-Kerr coupling strengths. The other parameters are fixed at $\Delta_0/\omega_b=0.5$, $\kappa/\omega_b=0.4$ and $P/P_0=2.0$.}
\end{flushleft}
\end{center}
\end{figure}

Finally, in Fig.\ref{fig6}, we plot $E_N$ as a function of the mean thermal phonon number $n_{th}$. We observe that the degree of entanglement monotonically decreases with increase in the thermal phonon number. However, the maximum number of thermal phonons up to which the entanglement persists increases with increase in the strength of the cross-Kerr coupling. For example, for the pure radiation-pressure coupling case the entanglement persists up to $n_{th}\approx460$. In contrast, in presence of a cross-Kerr coupling, with coupling strength $g_{ck}=10^{-3}g_0$, we have entanglement up to $n_{th}\approx4970$. This shows that with the introduction of cross-Kerr coupling, the bipartite entanglement becomes more robust against the thermal phonon fluctuations.It is worthwhile to mention that similar conclusions could be drawn for a different set of parameters as well. For example, we find that the proposed scheme works well for the following set of parameters, recently used in an experiment \cite{53}: $\omega_a/2\pi=195$ THz, $\omega_b/2\pi=3.68$ GHz, $g_0/2\pi=910$ KHz, $\gamma/2\pi=35$ KHz, $P_0=0.5$ mW and $n_{th}=100$. It is important to note that though with increase in the thermal phonon numbers entanglement gradually decreases, quantum correlation can persist upto to a much higher number of thermal phonons as quantified by the parameter, Gaussian quantum discord. This issue is addressed in the next section.

\section{\label{sec:level4}GAUSSIAN QUANTUM DISCORD}
For the bipartite system characterized by the correlation matrix $V$, Gaussian quantum discord can be expressed as \cite{54}:
\begin{align}\label{eq:17}
\mathscr{D}_G=f(\sqrt{I_1})-f(\nu^-)-f(\nu^+)+f(\sqrt{W}).
\end{align}
Here, the function $f$ is defined as:
\begin{align}\label{eq:18}
f(x)\equiv(x+\frac{1}{2})\ln(x+\frac{1}{2})-(x-\frac{1}{2})\ln(x-\frac{1}{2}),
\end{align}
\begin{align}\label{eq:19}
\nu^\pm\equiv2^{-1/2}\left[\Delta(V)\pm\sqrt{\Delta(V)^2-4I_4}\right]^{1/2},
\end{align}
are the two symplectic eigenvalues of $V$ with $\Delta(V)\equiv I_1+I_2+2I_3$, and
\begin{align}\label{eq:20}
W=\begin{cases}
    \left[\frac{2|I_3|+\sqrt{4I_3^2+(4I_1-1)(4I_4-I2)}}{(4I_1-1)}\right]^2 \quad \mathrm{if} \frac{4(I_1I_2-I_4)^2}{(I_2+4I_4)(1+4I_1)I_3^2}\leq1 \\\;
    \frac{I_1I_2+I_4-I_3^2-\sqrt{(I_1I_2+I_4-I_3^2)^2-4I_1I_2I_4}}{2I_1} \quad \mathrm{otherwise}.
\end{cases}
\end{align}
\begin {figure}[t]
\begin {center}
\includegraphics [height=4.5cm,width =7cm]{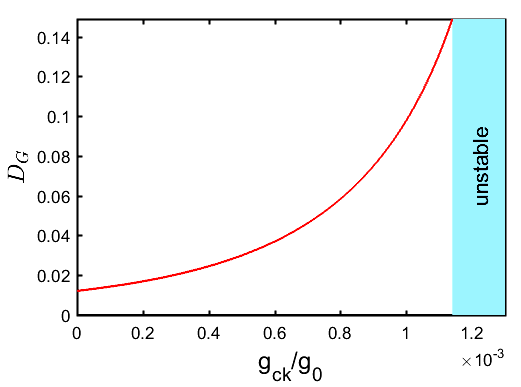}
\begin{flushleft}
\caption{\label {fig7}(Color online) $\mathscr{D}_G$ as a function of the normalized cross-Kerr coupling strength $g_{ck}/g_0$. The other parameters are fixed at $\Delta_0/\omega_b=0.5$, $\kappa/\omega_b=0.4$ and $P/P_0=2.0$.}
\end{flushleft}
\end{center}
\end{figure}
\begin {figure}[t]
\begin {center}
\includegraphics [height=4.5cm,width =7cm]{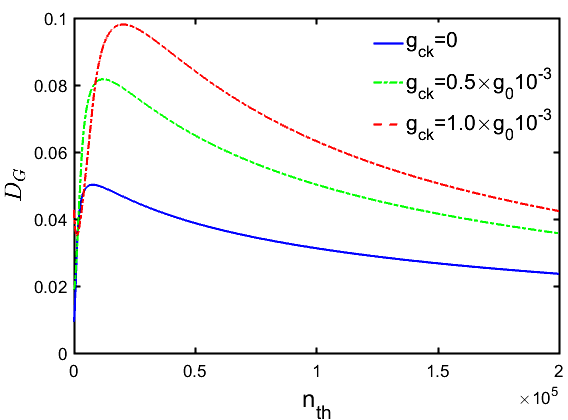}
\begin{flushleft}
\caption{\label {fig8}(Color online) $\mathscr{D}_G$ against mean thermal phonon number $n_{th}$. The other parameters are fixed at $\Delta_0/\omega_b=0.5$, $\kappa/\omega_b=0.4$ and $P/P_0=1.5$.}
\end{flushleft}
\end{center}
\end{figure}

To study the dependence of Gaussian quantum discord on the cross-Kerr nonlinearity, we first plot in Fig. \ref{fig7}, discord as a function of the normalized cross-Kerr coupling strength $g_{ck}/g_0$. We find that with increase in cross-Kerr coupling discord monotonically increases and becomes maximum just before the instability. For example, initially for the pure radiation-pressure coupling ($g_{ck}=0$) we have $\mathscr{D}_G\approx0.01$, on the other hand, in presence of an additional cross-Kerr coupling of coupling strength $g_{ck}=10^{-3}g_0$ we obtain $\mathscr{D}_G\approx0.1$. Therefore, this result along with Fig. \ref{fig3} confirms that incorporating cross-Kerr nonlinearity on a generic optomechanical system can considerably enhance the quantum correlation between the optical and the mechanical mode.

Next, in Fig. \ref{fig8} we plot the Gaussian quantum discord against the mean thermal phonon number, for increasing cross-Kerr coupling strengths. We observe that discord is a nonmonotonous function of the mean thermal phonon number. Discord increases initially for smaller number of thermal phonons but then, with increase in the thermal phonon numbers it decreases. However, it is worth mentioning that as compared to optomechanical entanglement, discord survives up to a significantly higher number of thermal phonons. We further note that at a fixed number of thermal phonon, increase in the cross-Kerr coupling strength enhances the degree of quantum discord. Therefore, we infer that in the presence of an additional cross-Kerr coupling the quantum correlation present in the system become more robust against the thermal phonon fluctuations.

\section{CONCLUSION}
In conclusion,we present a scheme to enhance steady-state entanglement between the optical and mechanical mode at a considerably lower driving power, by incorporating an additional cross-Kerr type coupling between the optical and mechanical mode. In addition to that, we show that with judicious choice of the cross-Kerr and the radiation-pressure coupling strength, it is possible to enhance quantum quantum correlation between the optical and the mechanical mode in an optomechanical system.It is shown that owing to cross-Kerr coupling, the system becomes unstable at  a considerably lower driving power resulting in significant enhancement of steady-state optomechanical entanglement near the instability threshold. Moreover, the bipartite entanglement becomes more robust against the variation of the system parameters and thermal phonon fluctuations with increasing cross-Kerr coupling strength.Further, we show that the quantum correlation which is quantified by the parameter called, quantum discord, unlike quantum entanglement, could persists upto a much higher bath temperature in the presence of cross-Kerr coupling. This work paves the way for exploring cross-Kerr coupling as a promising way to optimise both quantum entanglement and quantum correlation in a generic optomechanical system.

\section*{ACKNOWLEDGEMENT}
S. Chakraborty would like to acknowledge the financial support from MHRD, Government of India.

\end{document}